\documentclass{aa}  
\usepackage{amssymb,times,graphicx}
\usepackage[english]{babel}
\usepackage[varg]{txfonts}
\usepackage{natbib,afterpage}
\bibpunct{(}{)}{;}{a}{,}{,}
\usepackage{pgffor}
\usepackage{longtable}
\usepackage{amssymb,times,graphicx, subfigure}
\usepackage[english]{babel}
\usepackage{txfonts}
\usepackage{xspace}
\usepackage{longtable,lscape}
\usepackage{natbib,afterpage}
\usepackage{rotating}
\bibpunct{(}{)}{;}{a}{,}{,}

\def\iktau{IK~Tau\xspace}
\def\vycma{VY~CMa\xspace}
\def\cit6{CIT~6\xspace}

\def\rdor{R~Dor\xspace}
\def\rcas{R~Cas\xspace}
\def\txcam{TX~Cam\xspace}
\def\wxpsc{WX~Psc\xspace}
\def\waql{W~Aql\xspace}
\def\whya{W~Hya\xspace}
\def\chicyg{$\chi$~Cyg\xspace}
\def\omicet{o~Cet\xspace}
\def\raqr{R~Aqr\xspace}

\def\msunyr{$M_{\sun}$\,yr$^{-1}$\xspace}
\def\kms{km\,s$^{-1}$\xspace}

\def\alumina{Al$_2$O$_3$\xspace}
\def\irames{IRAM~30\,m\xspace}
\def\fwhm{$FWHM$\xspace}

\usepackage{amstext}

\begin{document}

\title{
Search for aluminium monoxide in the winds of oxygen-rich AGB stars\thanks{{\it Herschel} is an ESA space observatory with science instruments provided by European-led Principal Investigator consortia and with important participation from NASA.}\fnmsep\thanks{This work is partially based on observations carried out with the IRAM 30m Telescope. IRAM is supported by INSU/CNRS (France), MPG (Germany) and IGN (Spain).}
}

   \author{E. De Beck\inst{1}
          \and L. Decin\inst{2,3}
                \and S. Ramstedt\inst{4}
                \and H. Olofsson\inst{1}
          \and K. M. Menten\inst{5}
          \and N. A. Patel\inst{6}
          \and W. H. T. Vlemmings\inst{1}
          }

   \institute{Department for Earth and Space Sciences, Chalmers University of Technology, Onsala Space Observatory, Sweden
                  \\ \email{elvire.debeck@chalmers.se}
              \and Instituut voor Sterrenkunde, Departement Natuurkunde en Sterrenkunde, Celestijnenlaan 200D, 3001 Heverlee, Belgium
              \and Sterrenkundig Instituut Anton Pannekoek, University of Amsterdam, Science Park 904, 1098 Amsterdam, The Netherlands 
              \and Department of Physics and Astronomy, Uppsala University, Box 516, 75120 Uppsala, Sweden
                        \and Max-Planck-Institut f\"ur Radioastronomie, Auf dem H\"ugel 69, 53121 Bonn, Germany
              \and Harvard-Smithsonian Center for Astrophysics, 60 Garden Street, MS78, Cambridge, MA 02138, USA
             }

   \date{Received --- ; accepted ---}

  \abstract
   {Aluminium monoxide, AlO, is likely efficiently depleted from the gas around oxygen-rich evolved stars to form alumina (\alumina) clusters and dust seeds. The presence of AlO gas in the extended atmospheres of evolved stars has been derived from optical spectroscopy. More recently, AlO gas was also detected at long wavelengths around the supergiant \vycma and the oxygen-rich asymptotic giant branch star \omicet (Mira~A). The central role aluminium might play in dust formation and wind driving, in combination with these first detections of AlO at long wavelengths, shows the need for a wider search for this molecule in the winds of evolved stars.}
   {The detection at long wavelengths of emission in rotational transitions of AlO towards asymptotic giant branch stars can help constrain the presence and location of AlO gas in the outflows and ultimately also the efficiency of the depletion process. }
   {In search of AlO, we mined data obtained with APEX, the IRAM~30\,m telescope, \emph{Herschel}/HIFI, SMA, and ALMA, which were primarily aimed at studying other species around AGB stars. We report here on observations of AlO towards a sample of eight oxygen-rich asymptotic giant branch stars in different rotational transitions, up to seven for some stars.
   }
   {We present definite detections of one rotational transition of AlO for o Cet and R Aqr, and tentative detections of one transition for R Dor and o Cet and two transitions for IK Tau and W Hya. The presented spectra of \wxpsc, \rcas, and \txcam show no signature of AlO. For \omicet, \raqr, and \iktau, we find that the AlO~($N=9-8$) emission likely traces the inner parts of the wind, out to only a few tens of AU, where the gas has not yet been accelerated to its terminal velocity. This is in agreement with recently published results from a detailed study on \omicet.}
   {The conclusive detections of AlO emission in the case of \omicet and \raqr confirm the presence of AlO in the gas phase in outflows of asymptotic giant branch stars. The tentative detections further support this. 
   Since most of the observations  presented in this study were obtained with stronger emission from other species than AlO in mind, observations with higher sensitivity in combination with high angular resolution will improve our understanding of the presence and behaviour of AlO.
      From the current data sets we cannot firmly conclude whether there is a direct correlation between the wind properties and the detection rate of AlO emission.  We hope that this study can serve as a stimulus to perform sample studies in search of AlO in oxygen-rich outflows.
}


\maketitle
\titlerunning{AlO in the winds of oxygen-rich AGB stars}
\authorrunning{E. De Beck et al.} 


\section{Introduction}\label{sect:introduction}
Dust and its driving role in the winds of oxygen-rich asymptotic giant branch (AGB) stars remain poorly characterised. Descriptions of both the size and the composition of the dust prove difficult, although they are essential in the description of the wind-driving mechanism \citep[e.g.][]{bladh2012}. The composition that is often assumed as standard when modelling the circumstellar environment (CSE) of oxygen-rich AGB stars  is dominated by small silicate grains \citep[e.g.][]{bladh2015,decin2010_iktau_nlte}. 

The presence of alumina (\alumina)  dust around oxygen-rich AGB stars was first suggested by \citet{onaka1989} and is now considered standard, with modest but non-negligible mass fractions  \citep[e.g. 8\% in the case of \whya,][]{khouri2015_whya}. Both theoretical and observational studies show that \alumina and Fe-free silicates can be formed very close to the star \citep[at $2-4$ stellar radii;][]{zhaogeisler2012,karovicova2013,dellagli2014_alumina,gobrecht2016_dustformation_iktau}, before Fe-containing silicates can be formed. However, owing to its high transparency, \alumina might not contribute strongly to the driving of the wind, unless it grows large enough to efficiently contribute through backward scattering of incident stellar light or serves as a seed that is coated with silicate dust, for instance,
concealing the original \alumina signature \citep[e.g.][]{karovicova2013,hoefner2016_al2o3}. 

Solid \alumina is thought to be formed through trimolecular dimerisation of the gas-phase species AlO \citep[aluminium monoxide,][]{gobrecht2016_dustformation_iktau}, although the latter authors do not exclude additional pathways involving gas-phase AlOH. 
While short-wavelength detections of AlO have been reported for multiple oxygen-rich Mira stars and OH/IR stars  \citep[e.g.][]{merrill1962_optical_mtype,banerjee2012}, the only detections of emission of purely rotational transitions of AlO (that is, at long wavelengths) in the circumstellar environment (CSE) of an oxygen-rich AGB star are those recently published for \omicet (Mira~A) by \citet{kaminski2016_omicet}.  These authors additionally also claim tentative detections of AlOH towards \omicet. Emission from AlO and AlOH in the far-infrared and  (sub)millimeter regime was otherwise only reported for the CSE of the red supergiant \vycma \citep{tenenbaum2009, alcolea2013_vycma_hifistars,kaminski2013_vycma_sma}. 

We present observations of rotational transitions of AlO for a sample of eight oxygen-rich AGB stars. Their mass-loss rates $\dot{M}$ range from $\sim10^{-7}$\,\msunyr up to a few $10^{-5}$\,\msunyr \citep[e.g.][]{debeck2010_comdot}, reflecting a variation in
the wind density of two orders of magnitude.  We describe the observations in Sect.~\ref{sect:observations} and  present and discuss the results in Sects.~\ref{sect:results} and \ref{sect:discussion}. We present our conclusions in Sect.~\ref{sect:conclusion}.


\begin{table*}[ht]\caption{Summary of AlO observations towards oxygen-rich AGB stars.}\label{tbl:obs}
\centering
\small
\begin{tabular}{p{0.9cm}crrrrrrccl}
\hline\hline\\[-2ex]
Source  & Transition    & Frequency\tablefootmark{a}    & $E_{\mathrm{up}}/k$   & $\delta\nu$\tablefootmark{b}    & Rms noise     & Peak & Int. &HPBW\tablefootmark{c} or& Telescope            & Observing date\\
                        & $N- N^{\prime}$       & (MHz) & (K)                   & (MHz)           & (Jy) & (Jy)   & (Jy\,\kms) & aperture (\arcsec) & &\\   
\hline\\[-2ex]
\raqr & $9-8$                           &344451.8               & 82.8            & 3.4                   &0.075\tablefootmark{d} & 0.014\tablefootmark{d}                          &0.134\tablefootmark{d}& $0.38\times0.32$ & ALMA & 21 May 2015\\
                        [1.5ex]
\txcam  &$29-28$                &1106980.2      & 798.1         & 10.0            &5.4            &                       --      \tablefootmark{e}               &--     \tablefootmark{e}& 19              & HIFI          &1 Apr 2011\\
                        [1.5ex]
\rcas&$29-28$                   &1106980.2      & 798.1         & 10.0            & 6.1           &                       --      \tablefootmark{e}               &--     \tablefootmark{e}& 19              & HIFI          &9 Jun 2010 \\
                        [1.5ex]

\omicet &$9-8$                  &344451.8               & 82.8          & 2.3                     &0.150\tablefootmark{d} &       0.126\tablefootmark{d}                  &2.128\tablefootmark{d}& $0.64\times0.42$                        & ALMA  & 25 Feb 2014\\
                                                                &&              & &&      &                               & &                     &&+ 3 May 2014 \\
                        &$29-28$                &1106980.2      & 798.1           & 7.5                   &6.7            &       22.0                                                                                    &138.2                  & 19              & HIFI          &20 Jul 2010\\
                        [1.5ex]
\rdor           &$5-4$                  &191399.2               &27.6                 &1.8                    &0.15   &       --      \tablefootmark{e}                                       &--     \tablefootmark{e}& 34                      &APEX & 22, 23 Nov 2015          \\
                        &$6-5$                  &229670.3               &38.7                 &2.4                    &0.17   &       --      \tablefootmark{e}                                       &--     \tablefootmark{e}&         29              &APEX & 15 Aug 2014              \\
                        &                                       &               & &&      &                               &                                                               &&&                     + 2 Dec 2014\\
                        &$7-6$                  &267936.6               &51.5           &2.9                    &0.20   &       --      \tablefootmark{e}                                       &--     \tablefootmark{e}&         26              &APEX & 13 Jun 2012              \\
                                                &               &               & &&      &                               & &                     &&+ 22 Nov 2012\\
                        &$8-7$                  &306197.3               &66.2           &3.3                    &0.21   &       --      \tablefootmark{e}                                       &--     \tablefootmark{e}&21                         &APEX & 17 Aug 2014              \\
                                                &               &               & &&      &                               & &                     &&+ 29 Nov 2014\\
                        &$9-8$                  &344451.8               &82.8           &3.7                    &0.16   &       --      \tablefootmark{e}                                       &--     \tablefootmark{e}&19                         &APEX & 11 Nov 2011              \\
                        &$29-28$                &1106980.2      & 798.1           & 7.5                   & 7.8           &       19.1                                                                                    &83.7& 19              & HIFI          &9 Jun 2010\\
                        [1.5ex]
\whya   &$20-19$                & 764603.1      & 385.6         & 7.5             & 3.5           &       8.5                                                                                     &98.2& 27                      & HIFI          &17 Jul 2010\\
                        &$29-28$                &1106980.2      & 798.1           & 7.5                   & 6.7           &       20.5                                                                                    &71.3& 19              & HIFI          &20 Jul 2010\\
                        [1.5ex]
\wxpsc  & $3-2$                 & 114845.9      & 11.1          & 4.0                     & 0.050 &       --      \tablefootmark{e}                               &--     \tablefootmark{e}       & 21                      &IRAM   & 17 Jan 2010\\
                        & $4-3$                 & 153124.1      & 18.5            & 4.0                   & 0.022 &       --      \tablefootmark{e}                               &--     \tablefootmark{e}       & 16                      &IRAM   & 17 Jan 2010\\
                        & $6-5$                 & 229670.3      & 38.7            & 2.9         & 0.110   &       --      \tablefootmark{e}                               &--     \tablefootmark{e}       & 28                      & APEX          & 22 Sep 2009\\
                        &$20-19$                & 764603.1      & 385.6           & 10.0          & 3.5           &       --      \tablefootmark{e}                               &--     \tablefootmark{e}       & 27                      & HIFI          &17 Jul 2010\\
                        &$29-28$                &1106980.2      & 798.1           & 10.0          & 6.1           &       --      \tablefootmark{e}                               &--     \tablefootmark{e}       & 19              & HIFI          &20 Jul 2010\\
                        [1.5ex]
\iktau  & $3-2$                         & 114845.9      & 11.1          & 4.0                     &       0.020&  --      \tablefootmark{e}                               &--     \tablefootmark{e}       & 21                      &IRAM   & 22 Jan 2010\\ 
                        & $4-3$                 & 153124.1      & 18.5            & 4.0                   &       0.010&  0.024                                                                   &1.66   & 16                      &IRAM   & 14 May 2009 \\
                        &                                       &               & &&      &                               & &&    &               + 22, 26, 27 \\
                        &                                       &               & &&      &                               & &&    &                                       Jan 2010 \\
                        & $8-7$                 & 306197.3      & 66.2            & 4.9                   &       0.030&  --      \tablefootmark{e}                               &       --      \tablefootmark{e}& $1\times1$&SMA  & 26 Jan 2010\\
                        & $9-8$                 & 344451.8      & 82.8            & 4.9                   &       0.026&  0.098                                                                   &       2.33                            & $1\times1$&SMA  & 21 Jan 2010\\
                        &$20-19$                & 764603.1      &385.6          & 7.5                     & 2.5           &       --      \tablefootmark{e}                               &       --      \tablefootmark{e}&27                         &HIFI           & 7 Mar 2010\\
                        &$29-28$                &1106980.2      & 798.1           & 10.0          & 6.1           &       --      \tablefootmark{e}                               &       --      \tablefootmark{e}& 19              & HIFI          & 4 Mar 2010\\
\hline
\end{tabular}
\tablefoot{
\tablefoottext{a}{Weighted average of hyperfine-component frequencies, weighted with the component strengths.}
\tablefoottext{b}{Resolution at which the tabulated rms noise is measured. }
\tablefoottext{c}{Half-power beam width of the telescope main beam.}
\tablefoottext{d}{The rms noise and integrated intensity for the ALMA data are taken from the integrated-intensity maps. The appropriate units are hence also Jy\,beam$^{-1}$\,\kms. The peak values were taken from the spectra in Figs.~\ref{fig:omicet_alma} and \ref{fig:raqr_alma} and are in Jy\,beam$^{-1}$.}
\tablefoottext{e}{We do not list peak or integrated intensities for non-detections.}
}
\end{table*}

\section{Aluminium monoxide: AlO}
We report on observations of transitions between the rotational levels in the  $X^2\Sigma^+$  ground electronic state of the radical AlO. Owing to the nuclear spin of the Al atom, these transitions show resolvable hyperfine structure (HFS), which is specified in the entries for AlO in the JPL database for molecular spectroscopy \citep{lengsfield1982_alo_dipole,yamada1990_alo,pickett1998_jpl}. Throughout this report, we refer to the rotational transitions using the quantum number $N$, representing the rotational angular momentum. In this report we consider seven transitions $N=3-2
,
   4-3,6-5,8-7,9-8,20-19,\text{and } 29-28$, with upper-level energies $E_{\mathrm{up}}/k$ in the range $11-800$\,K.

\begin{figure}
\includegraphics[width=\linewidth]{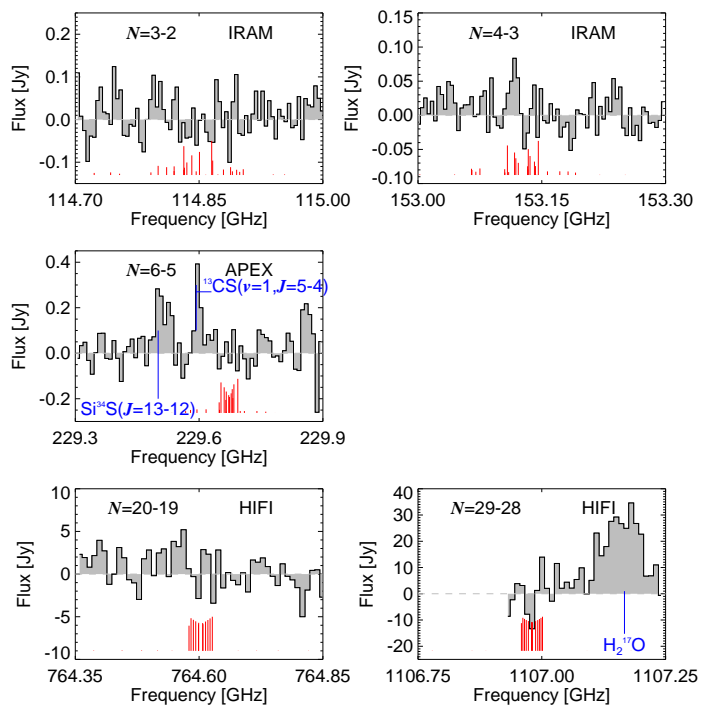}
\caption{Observations of AlO towards  \wxpsc: $N=3-2,4-3$ with \irames, $N=6-5$ with APEX, and $N=20-19,29-28$ with HIFI. The position and length of the red marks indicate the catalogue frequencies and relative strengths of the hyperfine structure components of each rotational transition.
\label{fig:wxpsc_alo}}
\end{figure}

\begin{figure}
\includegraphics[width=\linewidth]{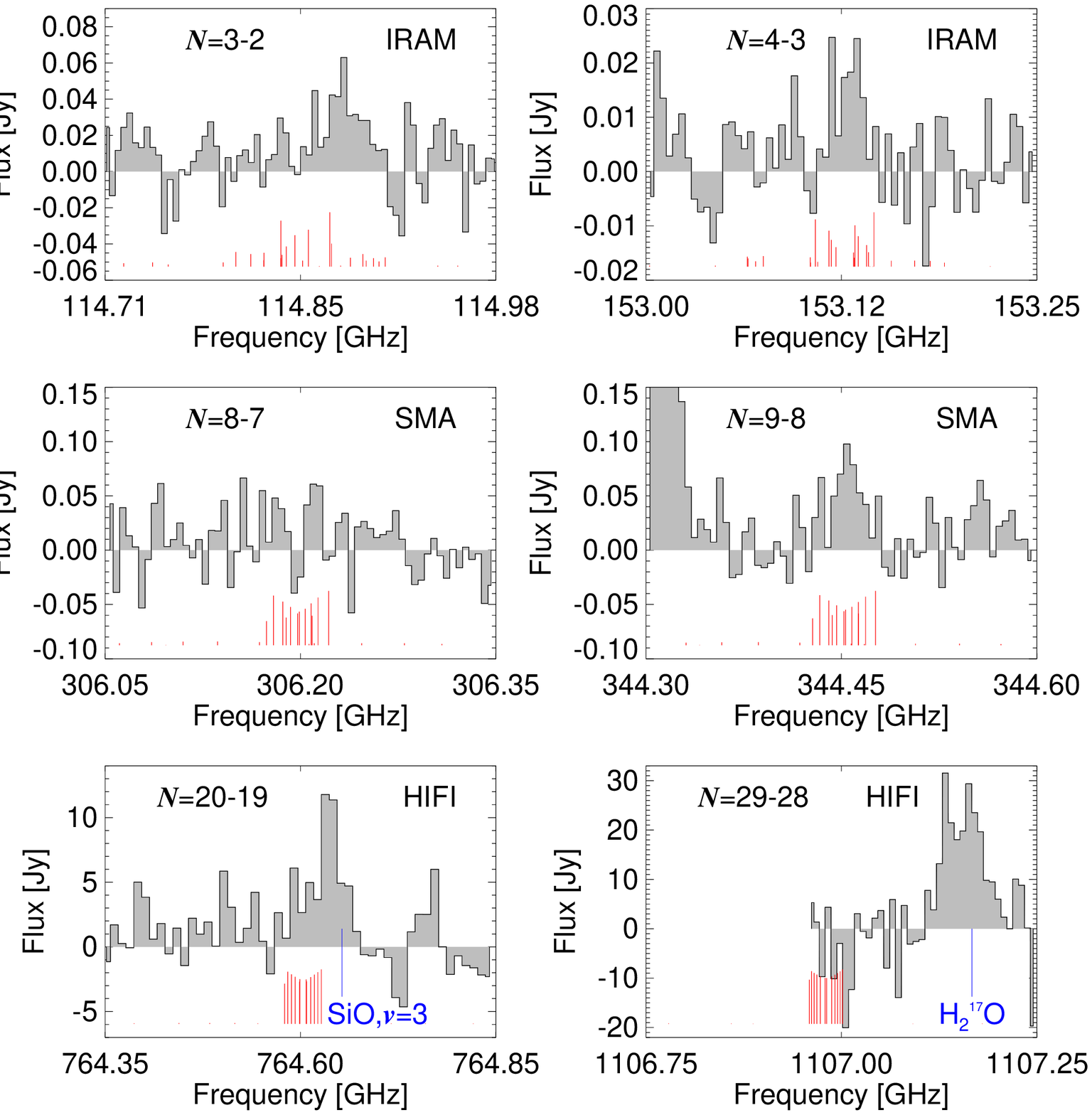}
\caption{Observations of AlO towards \iktau: $N=3-2,4-3$ with \irames, $N=8-7,9-8$ with SMA, and $N=20-19,29-28$ with HIFI. Red marks as described for Fig.~\ref{fig:wxpsc_alo}. \label{fig:iktau_alo}}
\end{figure}

\section{Observations}\label{sect:observations}
We note here that the observations are not homogeneous across the sample of eight stars, but rather stem from a collection of different observing programmes that we mined in search of AlO emission. These are projects that several of us were directly involved in, and we do not exclude the possibility that publicly accessible databases could harbour additional useful data for the presented or other AGB stars. These projects were not targeted at detecting AlO emission in particular, but we present several serendipitous detections and tentative detections below. Table~\ref{tbl:obs} and Figs.~\ref{fig:wxpsc_alo}\,--\,\ref{fig:raqr_alma} summarise the observations.

\subsection{\irames}
We observed the rotational transitions $N=3-2, 4-3$  of aluminium monoxide, AlO, at $\sim$114.8\,GHz and $\sim$153.1\,GHz towards two oxygen-rich stars, \iktau and \wxpsc, with the \irames telescope, using the EMIR single-pixel frontend, and the WILMA and 4M autocorrelator backends,  see Figs.~\ref{fig:wxpsc_alo} and \ref{fig:iktau_alo}. We present here the WILMA spectra, recorded at a nominal resolution of 2\,MHz, corresponding to 5.3\,\kms and 3.9\,\kms at the frequencies of the $N=3-2$ and $N=4-3$ transitions, respectively.  The half-power beam width (HPBW) of the main beam at these frequencies is 21\arcsec\/ and 16\arcsec, respectively. The observations towards both sources were carried out in beam-switching mode with a wobbler throw of 120\arcsec\/ to ensure stable spectral baselines and cancellation of contaminating emission along the line of sight. Data reduction was carried out using the \textsc{gildas/class}\footnote{\texttt{http://www.iram.fr/IRAMFR/GILDAS/}} package. We converted antenna temperatures into flux units by applying the point-source sensitivity as described on the web\footnote{\texttt{http://www.iram.es/IRAMES/mainWiki/Iram30mEfficiencies}}. 

In the case of \iktau, the $N=4-3$ transition was observed at two different epochs, 14 May 2009, and 22, 26, 27 January 2010, as part of two different projects. The earlier observations were part of a project to observe PO \citep[see][]{debeck2013_po_pn}, whereas the later observations targeted AlO specifically. We combine the data from both projects to increase the achieved sensitivity in the $N=4-3$ transition. 

\subsection{APEX}
We observed the $N=6-5$  transition at $\sim$229.6\,GHz towards \wxpsc (Fig.~\ref{fig:wxpsc_alo}) using the SHeFI heterodyne receiver \citep{vassilev2008_shefi} on APEX, at a nominal resolution of 0.5\,MHz, and with a HPBW of 28\arcsec.  The $N=5-4,\dots,9-8$ transitions were observed towards \rdor  (Fig.~\ref{fig:rdor_alo}) in the framework of a spectral survey with the SHeFI and SEPIA/Band-5 \citep{billade2012_band5} receivers on APEX, covering the range $159-368.5$\,GHz (De Beck et al., \emph{in prep.}). The observations were carried out in beam-switching mode with a wobbler throw of 50\arcsec\/ and 60\arcsec for \wxpsc and \rdor, respectively. 

Reduction of all APEX data was carried out using the \textsc{gildas/class} package. The conversion into flux units is based on point-source sensitivities of 38\,Jy/K, 39\,Jy/K, and 41\,Jy/K for the SEPIA/band-5 ($159-211$\,GHz), SHeFI-1 ($213-275$\,GHz), and SHeFI-2 ($267-378$\,GHz) observations, respectively.

\subsection{\emph{Herschel}/HIFI}
The $N=20-19$ and $N=29-28$ transitions of AlO at  764.603\,GHz and 1106.980\,GHz, respectively, were observed towards several evolved stars  using \emph{Herschel}/HIFI (henceforth HIFI) within the HIFISTARS guaranteed time key program \citep{bujarrabal2011_hifistars_success}. Detections of AlO based on these observations have, so far, only been reported for the red supergiant \vycma \citep{alcolea2013_vycma_hifistars}. The observations of the S-type AGB stars \chicyg and \waql do not show any AlO emission \citep{schoeier2011_chicyg,danilovich2014_waql}. \citet{justtanont2012_orich} presented the observations towards the oxygen-rich AGB stars in the sample, but did not indicate any AlO detections. However, we show here that there are some tentative detections of AlO emission in these HIFI observations. 

The HIFI observations were performed in beam-switching mode. HPBWs at the frequencies of the $N=20-19$ and $N=29-28$ transitions are 27.4\arcsec\/ and 19.1\arcsec\/, respectively. We retrieved the most recent level-2 products of the observations present in the Herschel Science Archive\footnote{\texttt{http://www.cosmos.esa.int/web/herschel/science-archive}}, processed with HIFI calibration tree  \texttt{HIFI\_CAL\_24\_0}. For a more detailed description of the observations we refer to the aforementioned papers.   The conversion of line intensities and rms noise levels into flux units is based on the point-source sensitivities reported by \citet{roelfsema2012_hifiinorbit}. 

\begin{figure}
\includegraphics[width=\linewidth]{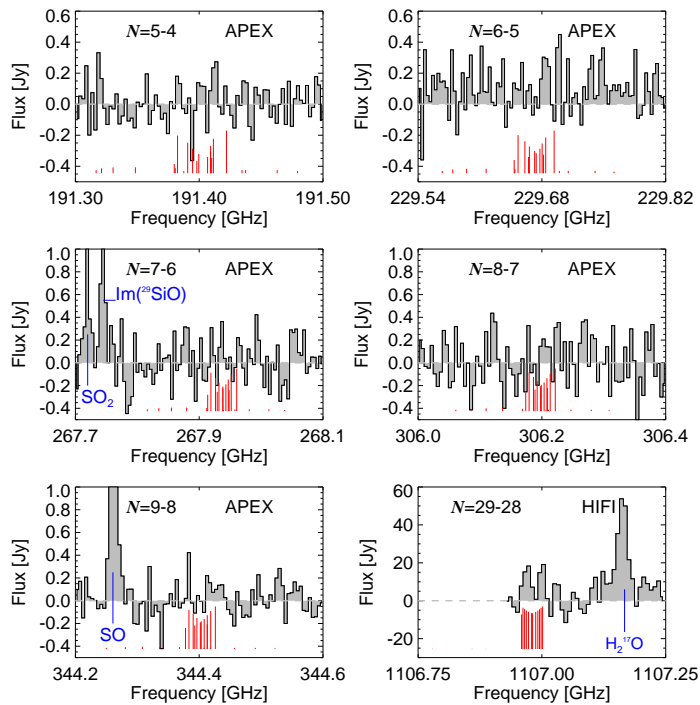}
\caption{Observations of AlO $N=5-4,\dots, 9-8$ towards \rdor with APEX and of $N=29-28$ with HIFI. Red marks as described for Fig.~\ref{fig:wxpsc_alo}. \label{fig:rdor_alo}}
\end{figure}

\begin{figure}
\includegraphics[width=\linewidth,trim={0 14cm 0 0},clip]{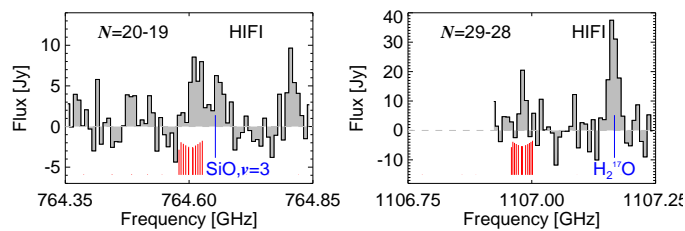}
\caption{Observations of AlO $N=20-19,29-28$ towards \whya with HIFI. Red marks as described for Fig.~\ref{fig:wxpsc_alo}. \label{fig:whya_alo}}
\end{figure}

\begin{figure}
\includegraphics[width=\linewidth,trim={0 7cm 0 0},clip]{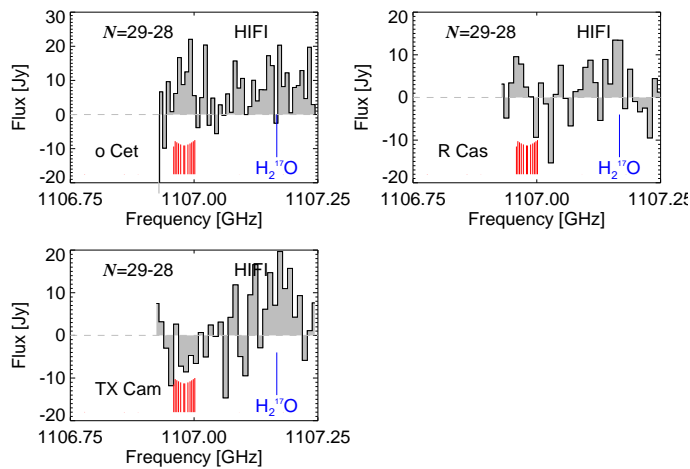}
\caption{Observations of AlO $N=29-28$ towards \omicet, \rcas, and \txcam with HIFI. Source names are marked in the lower left corner of each panel. Red marks as described for Fig.~\ref{fig:wxpsc_alo}. \label{fig:omicet_rcas_txcam_alo}}
\end{figure}

We show the spectra of the seven oxygen-rich AGB stars in the HIFISTARS sample in Figs.~\ref{fig:wxpsc_alo} -- \ref{fig:omicet_rcas_txcam_alo}. The $^{28}$SiO~($v=3,J=18-17$) transition at 764.653\,GHz is, although not identified by \citet{justtanont2012_orich}, possibly present in the spectra of oxygen-rich stars, and likely a maser. It is separated from the mean weighted frequency of the AlO~($N=20-19$) transition by only 20\,\kms (50\,MHz), which is less than or close to twice the outflow velocity of most of the observed stars. This makes blending of the lines highly likely in case they are both present in the spectrum. The proximity of the AlO~($N=29-28$) transition to the edge of the observed window makes detection of this line less reliable.

\subsection{SMA}
Within the framework of a spectral survey of \iktau in the range $279-355$\,GHz conducted with the Submillimeter Array (SMA), the $N=8-7,9-8$ transitions of AlO at $\sim$306.1\,GHz and $\sim$344.4\,GHz were observed.  For details on the observing strategy, calibration, data reduction, and overall characteristics of the observations, we refer to  \citet{debeck2013_po_pn} and \citet{kaminski2013_vycma_sma}. The SMA spectra we show in Fig.~\ref{fig:iktau_alo} are extracted for a square $1\arcsec\times1\arcsec$ aperture, large enough to cover the entire emitting region of AlO.

\subsection{ALMA}
\begin{figure}[ht]
\centering
\begin{minipage}{6.5cm}
\includegraphics[width=.7\linewidth,trim={0 0 1.5cm 0},clip]{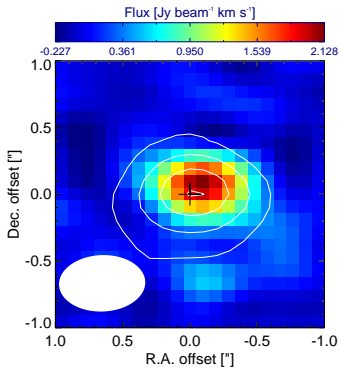}

(a) Colour map of integrated intensity. Contours indicate continuum emission at $3,10,20, \text{and }30$\,$\sigma$. The plus sign designates the position of the star, the filled white ellipse the synthesised beam.
\end{minipage}\hfill
\begin{minipage}{6.5cm}
\includegraphics[width=.7\linewidth]{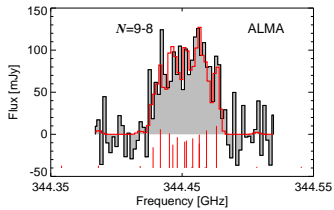}

(b) Spectrum extracted for the central pixel \emph{(black)}. The vertical red lines indicate the relative strength of the HFS components. The red histogram represents a multi-component synthetic line profile with a \fwhm of 4.5\,\kms.
\end{minipage}
\caption{AlO~($N=9-8$) emission towards \omicet, observed with ALMA. \label{fig:omicet_alma}}
\end{figure}

\begin{figure}[ht]
\centering
\begin{minipage}{6.5cm}
\includegraphics[width=.7\linewidth,trim={0 0 1.5cm 0},clip]{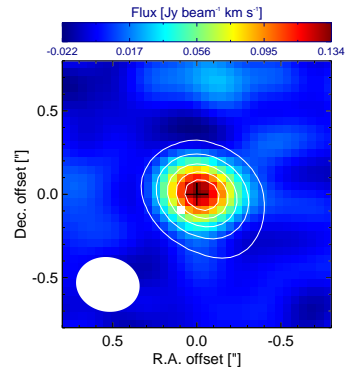}

(a) Colour map of integrated intensity. Contours indicate continuum emission at $3,10,20,30,\text{and }40$\,$\sigma$. The plus sign designates the position of the star, the filled white ellipse the synthesised beam.
\end{minipage}\hfill
\begin{minipage}{6.5cm}
\includegraphics[width=.7\linewidth]{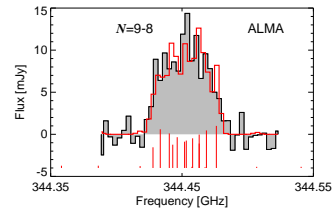}

(b) Spectrum extracted for the central pixel  \emph{(black)}. The vertical red lines indicate the relative strength of the HFS components. The red histogram represents a multi-component synthetic line profile with a \fwhm of 5.0\,\kms.
\end{minipage}
\caption{Same as Fig.~\ref{fig:omicet_alma}, but for \raqr. \label{fig:raqr_alma}}
\end{figure}

In the framework of a project aimed at studying the morphology of the outflows of a small sample of binary AGB stars, the oxygen-rich AGB stars \omicet and \raqr were observed with ALMA, the Atacama Large Millimeter/submillimeter Array, in emission of  CO~($J=3-2$). Some first results on \omicet and a description of the data calibration and reduction procedures have been presented by \citet{ramstedt2014_mira}. The spectral setup of the observations covers the AlO~($N=9-8$) transition at $\sim$344.4\,GHz.  The observations of \omicet have a synthesised beam size of $0\farcs64\times0\farcs42$, those of \raqr  $0\farcs38\times0\farcs32$. The integrated intensity maps in Figs.~\ref{fig:omicet_alma} and \ref{fig:raqr_alma} have rms noise levels of 0.15\,Jy\,beam$^{-1}$ (\omicet) and 0.075\,Jy\,beam$^{-1}$ (\raqr).

We note that the AlO spectra and maps presented by \citet{kaminski2016_omicet} for \omicet, which show clear detections of the $N=6-5,9-8$ lines, involve observations obtained within other observing projects and are hence not identical to the data shown in Fig.~\ref{fig:omicet_alma}.

\section{Results}\label{sect:results}
\subsection{Overview of the sample}\label{sect:overview}
\textbf{\raqr} is a Mira pulsator at a distance $d=214$\,pc, with a mass-loss rate of $\sim 0.5-1.0\times10^{-5}$\,\msunyr, and is the AGB primary star of a symbiotic binary system \citep{bujarrabal2010_symbiotic}. We detect emission from $N=9-8$ with ALMA and show the integrated intensity map and spectrum in Fig.~\ref{fig:raqr_alma}. The emission reaches a signal-to-noise ratio $S/N\approx18$ in integrated intensity, and the spectrum shows a line profile that is appropriately broadened by the HFS of AlO to a full-width at zero-power of $\sim 60$\,\kms, while CO emission arises from a region with wind velocities in the range $5-25$\,\kms \citep{bujarrabal2010_symbiotic}. From a comparison to the continuum emission, we conclude that the AlO emission coincides with the stellar position. The current spatial resolution ($0\farcs38\times0\farcs32$) does not allow us to resolve the AlO emission, but assuming a distance of 214\,pc, we can constrain the emitting region to the inner $\sim35$\,AU (in radius) around the star.

\textbf{\txcam} and \textbf{\rcas} are Mira pulsators with mass-loss rates of $\sim4.0\times10^{-6}$\,\msunyr and $\sim 8\times10^{-7}$\,\msunyr, at distances $d=380$\,pc and $d=176$\,pc, respectively \citep[e.g.][]{maercker2016_water}. We do not detect emission from the  $N=29-28$ transition in the HIFI spectrum of either \txcam or \rcas (Fig.~\ref{fig:omicet_rcas_txcam_alo}).  This is in line with their generally lower observed line intensities compared to the other stars in the sample \citep{justtanont2012_orich}.  

\textbf{\omicet}, Mira~A, is the AGB primary star of the well-known Mira binary  system. \omicet is nearby ($d=92$\,pc) and has a mass-loss rate of $\sim10^{-7}$\,\msunyr \citep[][and references therein]{ramstedt2014_mira}. We report a definite detection of the $N=9-8$ transition with ALMA (Fig.~\ref{fig:omicet_alma}) at a $S/N\approx14$ in integrated intensity. The extracted spectrum shows a clearly broadened line profile. Whereas the wind of \omicet has an expansion velocity of only $\sim5$\,\kms \citep{ramstedt2014_mira}, we see a full-width at zero-power of $\sim 60$\,\kms. We discuss the velocity of the AlO gas in more detail in Sect.~\ref{sect:velocity}. The emission is spatially unresolved, but we constrain the emitting region to the inner $20$\,AU (in radius) around the star. We further claim a potential tentative detection of the $N=29-28$ transition with HIFI at  $S/N\approx3.3$ (Fig.~\ref{fig:omicet_rcas_txcam_alo}), in line with the value reported by \citet{kaminski2016_omicet}. We note here that all values of $S/N$ that do not relate to ALMA data refer to peak $S/N$s. The uncertainties on this quantity are smaller than those on $S/N$ values based on integrated fluxes given the uncertainties on the boundaries of the emission lines in the spectra. A very detailed study of both short and long-wavelength observations of AlO is reported by \citet{kaminski2016_omicet}, who strongly constrain the presence of AlO around \omicet to the warmest parts of the CSE, out to only a few stellar radii, or $\sim10$\,AU.

\textbf{\rdor} is a nearby ($d=59$\,pc) semi-regular pulsator with a mass-loss rate of $\sim1.6\times10^{-7}$\,\msunyr \citep[e.g.][]{maercker2016_water}. The APEX and HIFI observations together cover six transitions of AlO towards \rdor: $N=5-4,\dots,9-8$ and $N=29-28$ --- see Fig.~\ref{fig:rdor_alo}. None of the transitions covered by APEX are detected at the current sensitivity. The $N=29-28$ transition is tentatively detected with HIFI, but the proximity of the emission to the band edge hampers a definite detection. We attempted to stack the spectra, given the high number of transitions covered, but the hyperfine structure components of the different AlO transitions do not align in frequency space, making stacking to confirm low-level detections less reliable. More observational support is needed to firmly claim the detection of AlO in the long-wavelength spectrum of \rdor. 

\textbf{\whya} is a nearby ($d=78$\,pc) semi-regular pulsator with a low mass-loss rate of $\sim1.0\times10^{-7}$\,\msunyr, with very similar basic properties to \rdor  \citep[e.g.][]{maercker2016_water}. We claim tentative detections of both transitions observed with HIFI, $N=20-19$ and $N=29-28$, at $S/N\approx2.4$ and $S/N\approx3.1$, respectively  (Fig.~\ref{fig:whya_alo}). Thanks to the very narrow line profiles in the spectrum of \whya, contamination or band edge effects are less important.

\textbf{\wxpsc}  is an oxygen-rich AGB star, often classified as an OH/IR star, with a very high mass-loss rate of a few $10^{-5}$\,\msunyr and at a distance $d\approx740$\,pc \citep[e.g.][]{debeck2010_comdot}. We show the  observed spectra in Fig.~\ref{fig:wxpsc_alo}. We report non-detections for the transitions $N=3-2, 4-3,\text{
and
} 6-5$ observed with APEX. We see a low-level peak at a $S/N\approx3.8$ in the frequency range of the $N=4-3$ transition, but find no agreement with the intrinsic HFS of the transition, concluding that this transition is not detected. The HIFI spectra  of $N=20-19$ and $N=29-28$  show no detectable emission at rms noise levels of 3.5\,Jy and 6.1\,Jy, respectively, at a resolution of 10\,MHz.

\textbf{\iktau} is a relatively nearby ($d\approx265$\,pc) Mira pulsator with a high mass-loss rate of $\sim8\times10^{-6}$\,\msunyr \citep[e.g.][]{decin2010_iktau_nlte}. We do not detect emission from the $N=3-2$ and $N=8-7$ transition at the current sensitivity of the observations, but report tentative detections of the $N=4-3
 \text{ and
} 9-8$ transitions of AlO with $S/N\approx2.4$ and  3.8 at rms noise levels of 10\,mJy and 26\,mJy, and frequency resolutions of 4\,MHz and 4.9\,MHz, respectively.  While the HIFI observation of $N=29-28$ yields no detection close to the band edge, emission from the $N=20-19$ transition is possibly present in the HIFI spectrum; see Fig.~\ref{fig:iktau_alo}. However, judging from the shape of the detected emission, SiO~($v=3,J=18-17$) is possibly also excited in \iktau and is likely blended with the AlO emission, complicating an unambiguous detection. Furthermore, a detection of the $20-19$ transition would not be in line with the low excitation temperature we derive in Sect.~\ref{sect:iktau}. Based on this, we claim the tentative detection of AlO in the circumstellar envelope of \iktau. We present a more in-depth discussion on the AlO emission from \iktau in Sect.~\ref{sect:iktau}.

\subsection{Velocity of the AlO gas}\label{sect:velocity}
The hyperfine structure of the AlO rotational transitions, where the individual components are spaced by several MHz, causes the emission lines to be broadened compared to those of CO rotational transitions, for instance. Following the example of \citet{kaminski2013_vycma_sma,kaminski2016_omicet}, we can construct synthetic profiles that are the superposition of a Gaussian profile for each HFS component of the transition. Narrow Gaussian line profiles are thought to be emitted in the wind-acceleration zone \citep[][and references therein]{decin2010_iktau_nlte}. The strength of each component is scaled according to the theoretical line strengths found in the JPL database \citep{lengsfield1982_alo_dipole,yamada1990_alo,pickett1998_jpl}, which are representative of  optically thin radiation excited under LTE conditions. The same intrinsic line width, characterised by the full-width at half-maximum (\fwhm) of the Gaussian, is assumed for all hyperfine components of one $N- N^{\prime}$ rotational transition for each star. 

The sensitivity and frequency resolution of the ALMA observations of \omicet and \raqr allow us to consider details in the line shape that are a direct consequence of the HFS of AlO. From a visual inspection of the synthetic profiles produced for \fwhm in the range $0.5-25.0$\,\kms (in steps of 0.5\,\kms), we suggest best-fit values for \fwhm of $4.5\pm1.0$\,\kms and $5.0\pm1.0$\,\kms for \omicet and \raqr, respectively. These are based on spectra at a velocity resolution of 2.0\,\kms and 3.0\,\kms, respectively. See the bottom panels of Figs.~\ref{fig:omicet_alma} and \ref{fig:raqr_alma} for the comparison of the best-fit synthetic profiles to the ALMA observations of \omicet and \raqr.

Given the current sensitivity of the observations for \iktau,  \fwhm\  values of 5\,\kms, 15\,\kms, and 30\,\kms all seem to reasonably reproduce the observations, and we cannot select a best-fit model. The $N=9-8$ emission is reasonably reproduced with any of the tested \fwhm between 2.5\,\kms and 30\,\kms, although profile fits indicate an (uncertain) upper limit of $\sim6$\,\kms, and for a \fwhm exceeding 10\,\kms it might be argued that the synthetic profiles are slightly too wide. 

The best-fit \fwhm values appear to imply that the AlO $N=9-8$ emission originates from a region in the CSE where the gas has not been accelerated much past 5\,\kms. In the cases of \raqr and \iktau, with terminal wind-expansion velocities of $\sim15$\,\kms and $\sim18.5$\,\kms \citep[e.g.][]{bujarrabal2010_symbiotic,decin2010_iktau_nlte}, this result  constrains the emission region to the inner part of the CSE.  Based on the wind-velocity profiles presented by \citet{decin2010_iktau_nlte} and \citet{maercker2016_water} for \iktau, we can limit the outer edge of the emission region to $10-20$\,AU in radius. We note that this estimate is rather uncertain since the \fwhm is not very well constrained. The ALMA observations constrain the emission around \raqr to the inner 35\,AU (in radius) of the wind (see Sect.~\ref{sect:overview}). Based on the wind-velocity profile presented by \citet{bujarrabal2010_symbiotic}, a 5\,\kms velocity would further constrain this to the inner 8\,AU (in radius). Given the low terminal velocity of the wind of \omicet  \citep[$\sim5$\,\kms;][]{ramstedt2014_mira}, a \fwhm close to 5\,\kms does not strictly constrain the presence of AlO to the very inner wind regions. However, the ALMA emission maps analysed by  \citet{kaminski2016_omicet} do set such a constraint, with AlO only emitting out to $\sim10$\,AU from the star. We point out that the \fwhm we derive for \omicet is significantly lower than the $8-14$\,\kms derived by \citet{kaminski2016_omicet}, confirming the suggestion by the latter authors that the intrinsic broadening was indeed smaller in the 2014 observations.

\citet{gobrecht2016_dustformation_iktau} quote shock velocities of $10-32$\,\kms in the region of the CSE where AlO is present. Such high velocities could cause line profiles to have a significantly higher fitted \fwhm than what we currently find for \omicet and \raqr. However, the models of \citet{gobrecht2016_dustformation_iktau} are optimised to represent the inner CSE conditions of \iktau, and we can currently not strictly rule out high velocities in AlO for this source. Nevertheless, we do see a consistently poorer fit for values of \fwhm that exceed $\sim 6$\,\kms, leading us to speculate (as we did above) that, also in the case of \iktau, the AlO emission traces lower velocities. Additional high-resolution (both spectral and angular) and time-monitored observations are essential to constrain the velocity dispersion and its potential variability in the AlO gas.

\subsection{Time variability}\label{sect:iktau_var}
Throughout one pulsation period the physico-chemical conditions of the circumstellar material as well as the stellar radiation field can vary drastically \citep[e.g.][]{cherchneff2006,gobrecht2016_dustformation_iktau}, complicating the interpretation of possible variability in the line emission. Since AlO is expected only very close to the star based on the chemical model of \citet{gobrecht2016_dustformation_iktau}, which we likely confirmed in the previous section, these varying conditions could play an important role. In-depth studies will need to address these different factors by carefully monitoring AlO emission features from different stars over time, much like the work of \citet{kaminski2016_omicet} for \omicet. 

\iktau is the only star in our sample for which we have observations from different epochs of one given transition. We coincidentally observed the four transitions $N=3-2,4-3,8-7,\text{and }9-8$ of AlO quasi-simultaneously in January 2010, see Table~\ref{tbl:obs}.  However, as mentioned in Sect.~\ref{sect:observations}, the $N=4-3$ transition was observed at an additional (earlier) epoch in May 2009 as part of another project.  These two epochs are separated by $\sim$255~days, about half a pulsation period $P$ for \iktau \citep[$P=470\pm5$ days;][]{wing1973_iktau_period_spectype}.   Based on the magnitude data available in the AAVSO database, we find that the May 2009 and January 2010 epochs correspond to intermediate (and similar) visual magnitudes and hence similar luminosities, but shortly after and shortly before a luminosity minimum, at respective phases $\phi\approx0.2$ and $\phi\approx0.7$. However, given the current sensitivity of the observations for \iktau, we cannot assess possible variability in the observed emission, and we perform any further analysis in Sect.~\ref{sect:iktau} on the spectrum that combines the two epochs. The HIFI spectra of \iktau  were obtained in the beginning of March 2010, that
is, at a maximum time difference of 45 days or a phase difference of $\sim0.1$, later than the bulk of the other spectra. We could safely ignore this 10\% difference in pulsational phase at this stage, but we choose to exclude the HIFI spectra of \iktau from the analysis below given the likely blend of the $N=20-19$ transition and the non-detection of the $N=29-28$ transition. Considering the location of the latter AlO transition in the observed window, we refrained from deriving an upper limit.

\begin{figure}[h]
\includegraphics[width=\linewidth,trim={0 14cm 0 0},clip]{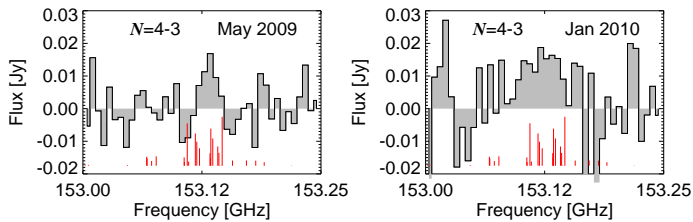}
\caption{\irames observations of AlO~($N=4-3$) towards \iktau from May 2009 \emph{(left)} and January 2010 \emph{(right)}. \label{fig:iktau_alo43_may2009_v_jan2010}}
\end{figure}

\subsection{AlO around \iktau}\label{sect:iktau}

\begin{figure}[h]
\includegraphics[width=\linewidth]{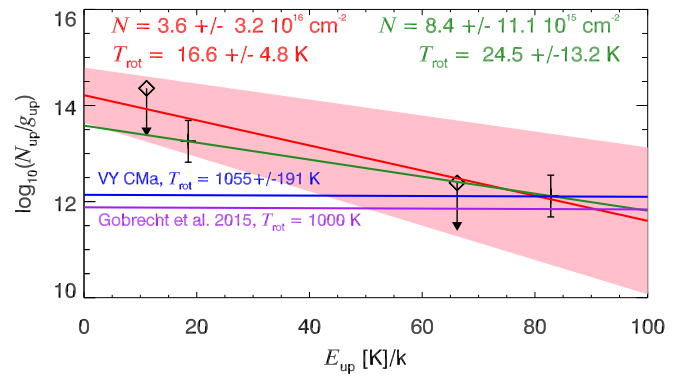}
\caption{Rotational temperature diagram for \iktau, assuming an emitting region of 75\,mas (20\,AU in radius). The red line and red shaded region indicate the fit to the data and 1$\sigma$ errors on the fit parameters.  The green fit is based only on the $N=4-3 \text{ and }9-8$ transitions. The purple line represents an abundance of $2.6\times10^{-8}$ in combination with a likely representative rotational temperature of 1000\,K \protect\citep{gobrecht2016_dustformation_iktau}. The blue line represents the $T_{\mathrm{rot}}$ and $N_{\mathrm{col}}$ values found by \protect\citet{kaminski2013_vycma_sma} for AlO around \vycma. \label{fig:AlO_rotdiagram} }
\end{figure}

Based on the set of observations we present here, \iktau is the only star in our sample for which we can attempt an abundance analysis. To derive an approximate abundance for AlO in the inner wind of  \iktau, we perform a rotational-diagram analysis of the integrated intensities. Assuming the integrated intensities from the data or from the synthetic profiles (see above) gives consistent results. The results described below and shown in Fig.~\ref{fig:AlO_rotdiagram} are based on the observational integrated intensities of the ground-based observations.

Assuming optically thin emission that is excited under local
thermal equilibrium (LTE) conditions, and an emitting region of 75\,mas (in diameter; $\sim$20 AU in radius), we derive a column density $N_{\mathrm{col}}=3.6\pm3.2\times10^{16}$\,cm$^{-2}$. For an average outflow velocity of 5\,\kms and a constant mass-loss rate of $10^{-5}$\,\msunyr, this leads to a source-averaged abundance AlO/H$_2$ $\approx 9.5\pm8.4\times 10^{-8}$. \citet{gobrecht2016_dustformation_iktau} predict AlO/H$_2$ around \iktau to drop from $2.6\times10^{-8}$ at the stellar surface to $1.6\times10^{-10}$ at six stellar radii.  We derive a rotational temperature $T_{\mathrm{rot}}=16.6\pm4.8$\,K. For comparison, for \vycma \citet{kaminski2013_vycma_sma} find  $T_{\mathrm{rot}}=1055\pm191$\,K, and $N_{\mathrm{col}}$ of $1.9\pm0.1\times10^{16}$\,cm$^{-2}$. We include the results of \citet{kaminski2013_vycma_sma} and \citet{gobrecht2016_dustformation_iktau} in Fig.~\ref{fig:AlO_rotdiagram}. For the latter, we assumed a rotational temperature of 1000\,K. 

The value of $T_{\mathrm{rot}}$ we find is much lower than the kinetic temperature of the region where the AlO emission is assumed to originate ($500-2000$\,K). Assuming that the low value of $T_{\mathrm{rot}}$ is representative of the local physical conditions, the AlO gas is placed around \iktau in the outermost parts of the wind, which is not consistent with the low velocities we report in Sect.~\ref{sect:velocity}. We hence take the low $T_{\mathrm{rot}}$ as a strong indicator that the assumption of optically thin emission excited in LTE is incorrect and that the AlO around \iktau could be strongly subthermally excited. Since AlO is likely present in the shocked environment close to the star \citep{gobrecht2016_dustformation_iktau,kaminski2016_omicet}, the emission could trace highly non-LTE conditions, which would
invalidate the derived values for $T_{\mathrm{rot}}$ and $N_{\mathrm{col}}$ and possibly even introduce changes in the relative strengths of the hyperfine components. In case of sufficient $S/N,$  such "skewed" HFS strengths might even
be retrieved from the data, but this is unfortunately not possible for the available data. It is difficult to predict in which way $T_{\mathrm{rot}}$ and $N_{\mathrm{col}}$ would change when taking non-LTE conditions into account, and we should hence interpret our results with particular care. Full radiative transfer modelling of the emission would be needed to derive more realistic values, but here also the complex kinematics of the gas -- including turbulence, the possible combination of outward and inward motions of the gas, and high-velocity shocks -- will complicate the treatment of the radiative transfer. Given the current sensitivity of the observations, such modelling is far beyond the scope of this work.


\section{Discussion}\label{sect:discussion}
Aluminium monoxide is likely an essential ingredient in the formation of dust seeds around oxygen-rich evolved stars, both AGB stars and red supergiant stars, since it can condense out at high temperatures (see below). If its condensation, and hence its depletion from the gas phase, is efficient, AlO is expected to be present in its gaseous form only very close to the star, in regions where dust seeds and grains have not yet formed. Our observations confirm that the size of the AlO~($N=9-8$) emission region is likely limited to within a few tens of AU from the star in the cases of \omicet and \raqr, and possibly also in the case of \iktau.

The physical and chemical conditions in the innermost layers are directly affected by pulsation-induced shocks \citep[e.g.][]{cherchneff2006}. According to the models of \citet{gobrecht2016_dustformation_iktau}, which treat the wind of \iktau specifically, the AlO abundance drops by several orders of magnitude as shocks pass through the material. The timescale on which the abundance varies in these models  is directly linked to the pulsation period, typically a few hundred days for AGB stars. Such drastic changes in abundance can be translated into changes in the amount of emitting AlO gas, and could hence be noticeable as variability in the emission features throughout the pulsation cycle. The variability in the stellar radiation field and in the gas density, both linked to the stellar pulsations, could induce an additional form of variability in the line emission depending on the dominant excitation mechanism. The coupling of these different aspects of variability will need to be considered in future detailed analyses of AlO emission. In the case of \omicet, \citet{kaminski2016_omicet} show that there is indeed true variability in the AlO emission strengths, but that it is not directly correlated with pulsation phase. The authors were unfortunately unable to discern between changes in excitation temperature or column density as causes of the variability.

We currently lack either a high enough number of emission lines (e.g. \raqr) or sufficient sensitivity (e.g. \iktau) to reliably estimate abundances of AlO in these different outflows. Furthermore, a study of the temporal variability of this emission would be in place given the variability of the conditions in the part of the wind where the AlO gas is excited. 

Comparing the integrated fluxes of the observed $N=9-8$ lines of \omicet, \raqr,  and \iktau, correcting for distance, we find that the emission from \raqr is intrinsically weaker by a factor of $\sim2.5$ than that of \omicet, and that the emission from \iktau could be a similar factor $\sim2.5$ stronger than that of \omicet. However, to derive any kind of correlation between pulsational type or mass-loss rate (density) and AlO emission strength, we need conclusive (and monitored) detections of AlO emission for a larger sample of stars. This sample needs to cover a broad range of parameters such as pulsation period (and type), mass-loss rate, and wind expansion velocity. If similar mass-loss rates guaranteed similar intrinsic AlO strengths, then \rdor and \whya should also have detectable flux in the $N=9-8$ transition. In the case of \rdor, this would still be in agreement with the non-detection at the sensitivity of the presented APEX observations. 

From mid-infrared interferometric observations using VLTI/MIDI, \citet{zhaogeisler2012} and \citet{karovicova2013} showed the coexistence of the warm extended atmosphere and an \alumina dust shell, that is, at up to two stellar radii, for several oxygen-rich AGB stars. Both studies report on a warm silicate dust shell appearing only at larger distances of typically $\text{four to
eight}$ stellar radii. The separation of these two types of dust is supported by the chemical models of \citet{gobrecht2016_dustformation_iktau}. The latter predict condensation of \alumina to be possible in the dense post-shock gas just above the stellar surface at $1600-2000$\,K, whereas the silicates only exist at distances beyond 3.5 stellar radii. According to the same models, gaseous AlOH plays an active role in alumina dust cluster formation, and can only be formed as a result of shock chemistry. We note that no emission of AlOH is detected in the SMA survey of \iktau, consistent with the significantly lower intensities of AlOH compared to those of AlO in the survey of \vycma \citep{kaminski2013_vycma_sma}. No AlOH emission lines were covered by the spectral setup of the ALMA observations of \omicet and \raqr. Both the chemical models and the mid-IR observations suggest that any further studies of AlO and AlOH in the gas phase will have to rely on high-sensitivity observations at high angular resolution, as can be done with ALMA. The current frequency coverage of ALMA, $84 - 950$\,GHz, allows us to observe several of the AlO transitions $N= 3-2$ up to $N= 24-23 $, spanning a range of energies corresponding to $10-550$\,K. The recent work by \citet{kaminski2016_omicet} clearly shows the great value that also lies in studies of refractory species that correlate long- and short-wavelength spectroscopy. Additionally, studies of the detailed dust composition and distribution will be essential to understand the evolution of species such as AlO in CSEs.

\section{Conclusion}\label{sect:conclusion}
We presented a search for purely rotational line emission of aluminium monoxide, AlO, in the winds of eight oxygen-rich AGB stars: \raqr, \txcam, \rcas, \omicet, \rdor,  \whya, \wxpsc, and \iktau. The observations were obtained with several ground-based observatories (APEX, \irames, SMA, ALMA) and with HIFI on board the Herschel Space Observatory and cover a wide range of the rotational spectrum of AlO, from $N=3-2$ at 115\,GHz (10\,K above ground) up to $N=29-28$ at 1107\,GHz (800\,K above ground). 

We reported on conclusive detections for two stars (\omicet and \raqr), tentative detections in the winds of a further three of the sample stars (\iktau,  \rdor, and \whya), and non-detections for the remaining three (\wxpsc, \rcas and \txcam). HIFI observations show tentative detections of two transitions towards \whya at $S/N\approx3$ and one towards both \rdor and \omicet at $S/N=2.4$ and 3.3, respectively. For \iktau we presented observations of six transitions, with tentative detections at $S/N=2.4$ and 3.8 for two of them ($N=4-3,9-8$). 

The detections and non-detections towards both the low- and high-density (or mass-loss rate) objects indicate that the detectability of AlO is not necessarily directly linked to the density or mass-loss rate. However, our data set is far from  homogeneous; the presented data sets are a combination of spectra obtained in the framework of several different projects, most of them not aimed at detecting AlO emission. A deeper and more homogeneous search towards a well-defined sample is hence advisable. To draw firm conclusions on the presence or absence of AlO and its abundance, it is necessary to obtain targeted observations with better sensitivity than currently achieved for most stars, combined with a broad coverage of stellar and wind parameters. The combination of high angular resolution with extreme sensitivity will be a necessary key to detect the emission from AlO and to constrain the emitting region, which is likely limited to a few stellar radii.


\begin{acknowledgements}
The Submillimeter Array is a joint project between the Smithsonian Astrophysical Observatory and the Academia Sinica Institute of Astronomy and Astrophysics and is funded by the Smithsonian Institution and the Academia Sinica.\\
The Herschel spacecraft was designed, built, tested, and launched under a contract to ESA managed by the Herschel/Planck Project team by an industrial consortium under the overall responsibility of the prime contractor Thales Alenia Space (Cannes), and including Astrium (Friedrichshafen) responsible for the payload module and for system testing at spacecraft level, Thales Alenia Space (Turin) responsible for the service module, and Astrium (Toulouse) responsible for the telescope, with in excess of a hundred subcontractors.
\\
This paper makes use of the following ALMA data: JAO.ALMA\#2012.1.00524.S. ALMA is a partnership of ESO (representing its member states), NSF (USA) and NINS (Japan), together with NRC (Canada) and NSC and ASIAA (Taiwan), in cooperation with the Republic  of Chile.  The Joint ALMA Observatory  is operated by ESO, AUI/NRAO and NAOJ. 
\\
The APEX observations were obtained under project numbers E-084.D-0724A-2009, O-087.F-9319A-2011,  O-094.F-9318A-2014, O-096.F-9336A-2015.
\\
WV acknowledges support from Marie Curie Career Integration Grant 321691 and ERC consolidator grant 614264.
LD acknowledges support from the ERC consolidator grant 646758 AEROSOL and the FWO Research Project grant G024112N.
HO acknowledges financial support from the Swedish Research Council.
\end{acknowledgements}

\addcontentsline{toc}{chapter}{Bibliography}
\bibliographystyle{aa}
\bibliography{alo.bib}

\end{document}